\documentclass[letterpaper,twocolumn,10pt]{article}

\usepackage{usenix}
\usepackage{algorithm}
\usepackage{algpseudocode}
\usepackage{balance}
\usepackage{booktabs}
\usepackage{graphicx}
\usepackage{amssymb}
\usepackage{amsmath}
\usepackage{enumitem}
\usepackage{xspace}

\microtypecontext{spacing=nonfrench}

\providecommand{\Description}[1]{}

\newcommand{\basenameattacklong}{Web Cache Overflow}
\newcommand{\attacklong}{\basenameattacklong\xspace}

\newcommand{\basenameattack}{WCO}
\newcommand{\attack}{\textsf{\basenameattack}\xspace}

\newcommand{\nginx}{NGINX\xspace}

\newcommand{\basenamemetric}{TAR}
\newcommand{\metric}{\textsf{\basenamemetric}\xspace}

\begin{document}

\date{}

\title{Web Cache Overflow: Exploiting Imprecise Keys for Cache Degradation and Beyond}

\author{
{\rm Matteo Golinelli}\\
University of Trento
\and
{\rm Kaan Onarlioglu}\\
Akamai \& Northeastern University
\and
{\rm Bruno Crispo}\\
University of Trento
}

\maketitle

\begin{abstract}

Web caches support the scalability needs of contemporary web applications by storing frequently accessed objects closer to clients. Web caches are conceptually associative arrays, tracking stored objects using cache keys consisting of HTTP request fields. However, these cache keys are often imprecisely defined by website operators. This allows clients to craft a multitude of requests that target the same object, but map to different cache keys.

In this work, we show that request elements included unnecessarily in cache keys can be abused to create redundant cache entries. In susceptible deployments, sustained generation of such redundant copies reduces cache effectiveness and increases origin load, facilitating eviction-dependent attacks. Our experiments reproduce cache degradation across five stand-alone caching proxies and characterize how these parameters affect attacker cost and cache hit rate, potentially resulting in denial-of-service attacks. We conclude that precise cache-key design is the most direct mitigation against this abuse vector and should be recognized as a security best practice.

\end{abstract}

\section{Introduction}
\label{sec:introduction}

Web caches are reverse proxies that sit between clients and origin servers, storing copies of frequently accessed objects. Web caches have reached existential importance in meeting the low access latency requirements for clients and traffic offload demands for origins. Cache degradation can rapidly escalate to costly network congestion and denial-of-service (DoS) incidents. Web caches are ubiquitous in contemporary web architectures, and their efficient utilization is crucial.

More generally, the topic of caching has been a staple of computer science. Researchers have explored various caching strategies, cache replacement algorithms, and even adversarial techniques to impair cache efficacy or launch cache-based side-channel attacks for different hardware and software engineering contexts. However, web caches remain particularly susceptible to abuse due to two important factors: their immense exposure to \textit{untrusted input} that influences the caching decisions, and widespread web cache configuration anti-patterns that do not account for such potential abuse.

Cache stores can be characterized as an associative array abstract data type; they associate every cacheable object with a \textit{cache key}. Cache keys are n-tuples that may consist of any data element included in an HTTP request according to the website operator's needs, such as the request line components, header values, or structured data fields inside the body. Upon receiving a request that matches a caching rule, the web cache derives the corresponding cache key and looks up the object in the cache store using that key. If the object exists, the request is immediately served from the cache. Otherwise, if the cache key does not correspond to an existing cache store entry, the web cache forwards the request to the origin server, receives a fresh copy of the requested object in the corresponding response, and then caches it under the said key.

In practice, however, cache keys are often not \textit{precisely} defined. For example, a website operator who creates a caching rule for a static image object served from \texttt{"example.com/pic.jpg"} may simply define the cache key as the full URL, overlooking the possibility of query strings included in the URL. This oversight would result in a subsequent request for a functionally equivalent but syntactically different URL, such as \texttt{"example.com/pic.jpg?junk=123"}, yielding a different cache key, because the two URLs do not exactly match. If discovered, a client can then take advantage of this situation to craft a multitude of requests for the same object, but with different cache keys, causing the web cache to retrieve the object from the origin anew every time. This is an instance of a well-known trick called \textit{cache busting}, often used by web developers for testing application changes, without needing to purge caches to observe the effects of their changes.

Motivated by these observations that clients influence the HTTP requests used to derive cache keys, and that imprecise cache key definitions occur in practice, we hypothesize that attackers can systematically weaponize cache-busting techniques to create redundant cache entries. In susceptible deployments, sustained generation of such entries reduces cache effectiveness by forcefully evicting useful content. We call this attack \attacklong (\attack).

One obvious impact of \attack is increased origin load, which could lead to service degradation and escalate to a DoS. Moreover, since web caches are frequently deployed in shared cloud infrastructures and hosting providers, an attack resulting from one tenant's imprecise cache keys may DoS all tenants of the platform. However, attackers can also leverage the capability to purge popular cached objects for other nefarious ends. For example, \textit{cache poisoning} attacks that aim to trick caches into storing malicious content are on the rise. Such attacks are challenging to launch against frequently accessed objects (e.g., a JavaScript file embedded on the home page), as such objects are often perpetually cached due to their popularity. With \attack, an attacker could fill the cache to trigger purging of even popular objects, exposing a window of opportunity to poison that object's now unused cache key with malicious content.

A novel characteristic of \attack is that, unlike previous cache attacks in the literature, generating redundant entries does not require prior knowledge of the replacement algorithm or the popularity of legitimate objects. Instead, it is an attack enabled by the fundamental workings of HTTP and web caches; it can be exploited by any unauthenticated, unauthorized, unsophisticated Internet client, and it is difficult to mitigate without incurring high costs to website operators.

We test these claims through a detailed evaluation of \attack. We first reproduce cache degradation across five popular caching proxies, and then perform focused tests using \nginx to explore the influence of cache capacity and object size. Testing \attack against production websites is not ethically feasible, as DoS attacks are disallowed in bug bounty programs. In lieu of penetration testing, we conduct a measurement study on the Tranco top 10k domain list to identify equivalent objects cached under different cache keys, the prerequisite condition for the attack.

We finally discuss seemingly intuitive mitigations, namely cache store quotas, request rate limiting, and object deduplication, including their deployment tradeoffs. Our experiments show that, in realistic scenarios, these mitigations may be ineffective, counter-productive, or unreasonably costly. We conclude that addressing the prerequisite condition by defining \textit{precise} cache keys is the most direct defense, and we document this as a security best practice for the first time.

To summarize, we make the following contributions.

\begin{itemize}
    \item We show for the first time that imprecise cache keys can be abused to create redundant cache entries and degrade cache effectiveness in susceptible web cache deployments.
    \item We present \attack, a cache pollution attack that increases eviction pressure on legitimate cached objects by sustaining redundant entries.
    \item We evaluate the parameters that influence \attack's feasibility and efficacy, characterizing its origin-load consequences and other potential implications, such as assisting cache-poisoning attacks.
    \item We present possible mitigations and evaluate them, concluding that defining precise cache keys is the preferred method rather than workarounds.
    \item We release an open-source penetration testing tool to assist operators in automatically detecting imprecise cache keys on their websites.
\end{itemize}

\paragraph{Availability}

We release two open-source tools: a penetration testing utility for detecting imprecise cache keys, intended for website operators and security testers, and an implementation of our attack to evaluate systems against \attack. All the code, data, and artifacts required to reproduce our experiments are also available in our repository at \url{https://github.com/Golim/web-cache-overflow}. The repository includes several README files with details about the repository contents and instructions for reproducing our experiments.

\section{Background \& Related Work}
\label{sec:background}

\subsection{Web Caches}

Websites are rapidly growing in size due to an increasing number of objects necessary to render each page, such as images, videos, fonts, JavaScript files, and style sheets~\cite{almanac}. Furthermore, latency-sensitive media streaming services and bandwidth-hungry big file downloads have become commonplace. All of these can put an enormous strain on an origin server's network resources if left unaddressed. Therefore, web caches have become an essential element of the modern web.

Web caches are positioned between a client and an origin server to temporarily store frequently accessed objects, preventing unnecessary repeated data transmissions and reducing both bandwidth and latency. Multiple caches can be present on the delivery path, starting with client-side caches (e.g., implemented inside a web browser), layers of reverse proxy servers acting as intermediary caches, and finally caches co-located with the origin server, forming cache hierarchies. Content Delivery Networks (CDNs) that provide performance and security services over distributed reverse proxy networks similarly cache content as a core capability.

In this paper, we use the term \textit{web cache}, or simply \textit{cache}, to refer to server-side web caches. We leave client-side caches out of scope, as attacks on such caches (i.e., a client attacking itself) do not constitute a valid threat model. We also do not explore niche uses of private caches, but instead focus our discussion on the far more prevalent public caches that have become a crucial part of web architectures.

\subsection{Cache Control}

HTTP provides a standard mechanism for origin servers to signal to the caches on the path whether a response may be stored, via the \texttt{Cache-Control} response header. The directives that can be included with this header define the cacheability of an object, its expiration time, and the revalidation actions.

RFC 9111 states that caches \textit{MUST} respect \texttt{Cache-Control} directives. However, this is not the case in practice, as also documented by prior academic work (e.g.,~\cite{mirheidari2020cached}). Website structures and caching policies may frequently change, which makes maintenance of \texttt{Cache-Control} directives to reflect those changes an operational burden for website operators, especially in enterprise infrastructures. Therefore, popular cache technologies offer configuration options to disregard \texttt{Cache-Control} headers, but instead implement \textit{cache rules} centrally at the cache. Cache rules can be made as general or specific as needed, allowing website operators to craft them via domain-specific languages or regular expression matches on virtually any part of a request or its corresponding response (e.g.,~\cite{cloudflare_cacheoverride,varnish_vcl}).

Once a response is determined to be cacheable, the cache derives a \textit{cache key} for the respective stored object. The cache key acts as a unique index into the cache store, enabling the cache to easily check whether it already holds a copy of the object upon receiving subsequent requests for the same resource. That is, the cache store implements an associative array as an abstract data type, keyed with the cache key.

Cache keys are n-tuples (or hashes of those n-tuples) that may consist of any element of an HTTP request, including the method, path, query string parameters, headers and their values, or structured data fields in the body payload. Request elements that are included in the cache key are called \textit{keyed}, and the rest \textit{unkeyed}. A typical default cache key is a simple 2-tuple consisting of the full URL (e.g.,~\cite{varnish_cachekey}), but additional elements may also be keyed for more complex situations, for instance, the \texttt{Origin} header for CORS support (e.g.,~\cite{cloudflare_cachekey}). Website operators customize cache keys for their needs.

A \textit{cache purge} is the process of invalidating a cached object, meaning that the next request that matches that object's cache key will be a cache miss. That request will then be forwarded to the origin server, and the cache will store the new response under the same cache key. Web caches provide secure mechanisms for website operators to purge caches, typically through an authenticated web interface or an API. Arbitrary web clients should never be able to purge the cache, as this would facilitate DoS attacks, cache poisoning, and cache-timing-based side-channels.

\subsection{Cache Busting}
\textit{Cache busting} collectively refers to client-side techniques used for intentionally bypassing a cache and receiving a fresh copy of the requested object from the origin server. Cache busting is frequently used in web development to rapidly test website changes without needing to purge any intermediary caches (which may not even be under the developers' control) that may still hold stale copies of modified resources. Similarly, penetration testers often leverage cache busting to avoid poisoning cacheable endpoints with attack payloads.

According to RFC 9111, clients may attempt cache busting by including a \texttt{Cache-Control} header set to the ``no-cache'' directive in their requests~\cite{rfc9111}. However, caches are not required to honor this request directive, and many production infrastructures in fact do not, due to the aforementioned security implications. Instead, practical cache busting techniques involve identifying keyed elements of a request that do not alter the response, and intentionally modifying those elements in requests to trigger a cache miss.

\subsection{Cache Status Headers}

Web caches often add cache status headers to responses, indicating how they handled the corresponding request. This field may simply include a cache hit versus miss flag, or more detailed debugging information.

RFC 9211 aims to standardize how caches communicate this information, but sadly, it has not yet been widely adopted~\cite{rfc9211}. Instead, each web cache technology uses its own custom headers to this end, and these are not always officially documented. Mirheidari et al., however, compiled a list of the different header names and values observed when interacting with a selection of popular stand-alone cache servers and CDNs~\cite{mirheidari2022web}.

\subsection{Cache Pollution}

Achieving good cached data locality through optimizing access patterns, adopting the appropriate cache eviction strategies, and designing cache hierarchies have been studied in computer science for decades. There is an immense body of literature on this topic as it applies to software, hardware, and network engineering domains. These works define the term \textit{cache pollution} to broadly refer to conditions that lead to a disruption of the cached data locality, resulting in a decrease in cache hits, and therefore performance degradation. Cache pollution is not necessarily an adversarial event, but rather, an outcome of less-than-optimal cache design given the use case. As an early but representative example relevant for a web application context, researchers showed that automated access patterns of web robots significantly decreased cache locality for popular objects~\cite{almeida2001analyzing}.

Cache pollution can also be induced for malicious purposes, aptly called a \textit{cache pollution attack}. Core attack concepts generally apply to all caches, and the adversarial techniques presented in all works can be summarized in two categories: 1) Attacks that repeatedly request the same unpopular objects, and 2) attacks that request a wide array of objects. Both techniques hurt the locality of genuinely popular objects. Defenses fall into various anomaly detection approaches based on access characteristics and patterns, and specialized cache replacement algorithms to minimize attacker influence~\cite{edgedefense}.

For network applications of caches, the vast majority of literature is focused on Information-Centric Networking (ICN), which encompasses Content-Centric Networking (CCN) and Named Data Networking (NDN). ICN is a novel network communications paradigm for the Internet that revolves around requesting named content, instead of host-to-host packet exchanges~\cite{rfc8793}. In-network caching plays a prominent role in this architecture, which resulted in a plethora of scholarly works exploring pollution attacks and defenses for the ICN paradigm (e.g.,~\cite{icn1,icn2,icn3,icn4,icn5,icn6,icn7,icn8,icn9,icn10}). We omit the specifics of individual works, as ICN is not relevant to our paper's scope. However, the attack and defense contributions are in line with the general summary we provided above.

Gao et al. and Deng et al. present incremental works more closely related to ours, exploring cache pollution on the web~\cite{gao2006internet,deng2008pollution}. These discuss the same two attack categories as applied to caching proxies and DNS caches, propose an anomaly detection scheme based on statistics computed over request features, and perform experiments using Squid. They primarily focus on pollution attacks in web and peer-to-peer cache deployments using forward proxies, as opposed to the \textit{reverse} proxies more prevalent today that we focus on. Nonetheless, the authors argue (and we agree) that the findings are also applicable to reverse proxies. More recently, Afek et al. explore a similar attack against DNS caches~\cite{afek2024flushing}. While the idea of flushing a cache via adversarial input is shared, the operational characteristics, defenses, and implications differ significantly between DNS and web caches.

In the next section, we formulate \attack, our novel approach to cache pollution. We defer discussion of the key differentiators between our work and the above literature until then.

\section{Research Statement}
\label{sec:research}

\begin{figure}[t]
    \centering
    \includegraphics[width=1.0\linewidth]{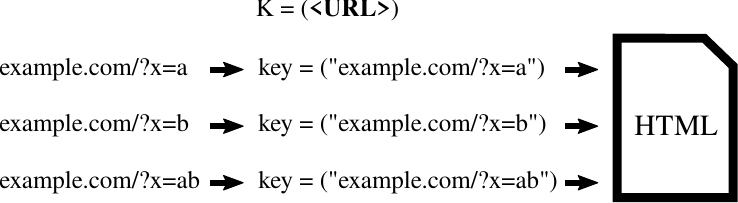}
    \caption{Cache pollution due to imprecise cache key definition, keyed on the full URL.}
    \label{fig:ex1}
    \Description{An example of cache pollution due to imprecise cache key definition, keyed on the full URL. The cache key is defined as the entire URL, including the host, path, and the complete query string. An attacker can attach arbitrary query strings that are not recognized by the application to repeatedly cache the homepage under different cache keys.}
\end{figure}

\subsection{The Problem}
The observation that motivates our research is that, while cache busting is often seen as a harmless and sometimes useful property of cache-enabled web architectures, the factors that enable cache busting can be used to generate redundant cache entries and increase eviction pressure.

We elaborate on the details of this observation and our resulting hypothesis below.

Let:
\begin{itemize}
    \item $K = (k_1, ..., k_n)$ be the n-tuple that represents the cache key definition, where $k_i \in K, i = 1,...,n$ describe the keyed HTTP request elements.
    \item $request_i$, $i \in \mathbb N$ be \textit{distinct} HTTP requests that result in cacheable responses containing payloads $object_i$.
    \item $key_i$ be the concrete cache keys derived from $request_i$ according to the definition $K$, resulting in the cache store mapping $key_i \rightarrow object_i$.
\end{itemize}

Then:
\begin{itemize}
    \item[] For any two $key_i \rightarrow object_i$ and $key_j \rightarrow object_j$,
    \item[] if $key_i \neq key_j$, but $object_i = object_j$,
    \item[] we say that $K$ is \textit{imprecise}.
\end{itemize}

\begin{figure}[t]
    \centering
    \includegraphics[width=1.0\linewidth]{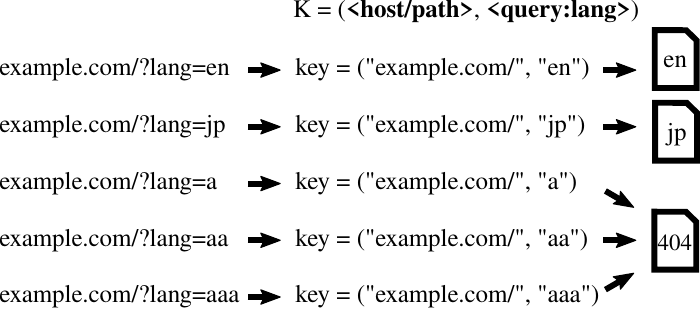}
    \caption{Cache pollution due to imprecise cache key definition, keyed on the query string parameter ``lang'', combined with insufficient validation of its value.}
    \label{fig:ex2}
    \Description{An example of cache pollution due to imprecise cache key definition, keyed on the query string parameter ``lang'', combined with insufficient validation of its value. The cache key is defined as the host, path, and a specific query string parameter that the application uses to serve a homepage translated into different languages. This mitigates the attack in the previous example, since arbitrary query strings do not influence the cache key. However, there is still room for abuse: If 1) the application does not validate the parameter against a list of supported languages, but allows any arbitrary string, and 2) the cache rules allow caching of 404 error pages--a common practice for traffic optimization--, then the attacker can still pollute the cache by repeatedly caching the same error page under different values for ``lang''.}
\end{figure}

Intuitively, a cache key definition that includes keyed request elements that do not influence the response payload results in the identical object getting stored multiple times, but under different cache keys, wasting space. This is a generalization and formalization of the basis of all cache busting, which may then be systematically abused for cache pollution. Web clients are free to craft and send any HTTP request to a server, and therefore, they can manipulate any keyed element included in $K$ in an attempt to achieve this effect.

The semantics of a \textit{precise} cache key definition naturally follow: Every unique concrete cache key derived from a precise definition maps to a unique cached object, eliminating this cache pollution vector.

Note that an imprecise cache key definition is a prerequisite, but not a sufficient condition for effective abuse. Whether cache pollution via imprecise cache key definitions is viable depends on request validation, which responses are cacheable, cache capacity, object-size distribution, and traffic rates.

Figure~\ref{fig:ex1} depicts an example where the imprecise cache key definition leaves little room for mitigating abuse through validation. Here, the cache key is defined as the entire URL, including the host, path, and the complete query string. As a result, an attacker can attach arbitrary query strings that are not recognized by the application to repeatedly cache the homepage under different cache keys.

Figure~\ref{fig:ex2}, on the other hand, illustrates a more complex situation. The cache key is now defined more prescriptively to only include the host, path, and a specific query string parameter that the application uses to serve a homepage translated into different languages. This mitigates the attack in the previous example, since arbitrary query strings do not influence the cache key. However, there is still room for abuse: If 1) the application does not validate the parameter against a list of supported languages, but allows any arbitrary string, and 2) the cache rules allow caching of 404 error pages--a common practice for traffic optimization--, then the attacker can still pollute the cache by repeatedly caching the same error page under different values for ``lang''.

Given the above observations, we present our hypothesis: An arbitrary Internet attacker can abuse imprecisely defined cache keys, keyed on insufficiently validated HTTP elements, to create redundant cache entries, launching pollution attacks. Sustained generation of such entries reduces cache effectiveness and may increase origin load or facilitate eviction-dependent attacks. We call this attack \attacklong (\attack).

\subsection{Impact and Novelty}

\attack fills a web cache with redundant copies of a specific object, forcefully purging everything else in the process. This negates the benefits of using a cache, drastically increasing the traffic load on the origin server and round-trip times for clients interacting with the application, ultimately resulting in a DoS similar to other cache pollution attacks.

However, \attack has a significant property that differentiates it from previous work: it requires no knowledge of the victim cache's regular access patterns, the popularity of objects served from the target website, or the cache eviction algorithm in use. Existing attacks rely on measurements and accurate estimations of such details for calculated, complex strategies that selectively boost the locality of unpopular files or disrupt the locality of popular ones, whereas such information is immaterial for \attack. The attack methodology requires no prior knowledge of the target environment and instead results in indiscriminate purging of objects, enabling even unsophisticated attackers. The sole requirement, identifying imprecise cache keys, can be trivially automated, as we discuss in Section~\ref{sec:method} and later demonstrate in Section~\ref{sec:large-scale-file-size}. Moreover, our work presents a fundamentally different attack vector and threat model compared to prior studies: previous attacks rely on pre-existing unpopular objects to fill the cache, inherently limiting their scale to the number and size of such objects available on the target site, whereas our approach employs cache-busting techniques to actively create arbitrary cache entries, removing this constraint. Thus, our attack is limited not by the site's content but only by factors such as available bandwidth, server-side rate limiting, the expiration age of cached entries, and the cache capacity. Finally, our methodology avoids the high cost of downloading large files by leveraging HEAD requests, further reducing the overhead of launching the attack.

Furthermore, \attack's ability to purge objects facilitates other attacks such as cache poisoning and cache-based side channels, which usually come with a pre-condition attached: The resource targeted with these attacks must not already be cached. Assume an attacker discovers a reflected cross-site scripting (XSS) vulnerability, where the application includes parts of the attacker-controlled request on the rendered page without input validation or output sanitization, allowing the attacker to inject a malicious JavaScript snippet into the page. If this poisonable page is also cacheable, an attacker may attempt to store the response containing the reflected XSS payload.
While such vulnerabilities are common, exploiting them is challenging if the page in question is getting a lot of traffic, implying that the original copy will perpetually be cached, not allowing the attacker to override it with a poisoned version. While \attack does not deterministically select the entry evicted by the replacement policy, sustained eviction pressure may create a window of opportunity for the target to become absent and for the attacker to poison the cache.

\subsection{Research Questions}
\label{sec:questions}
Our overarching goal is to test our hypothesis and characterize the conditions under which \attack degrades cache effectiveness by addressing the following research questions.

\begin{enumerate}[label=Q\arabic{enumi}]
    \item Are imprecise cache key definitions a common occurrence in real-life websites?
    \label{Q1}
    \item Under which deployment conditions is \attack practical? How does it affect cache effectiveness, and what traffic cost is required to maintain it?
    \label{Q2}
    \item How do various cache configurations, such as the total cache capacity, cached object size, web cache technology used, etc., influence the attack?
    \label{Q3}
    \item Can \attack be mitigated? How effective are the established cache defenses against \attack?
    \label{Q4}
\end{enumerate}

\subsection{Threat Model}

We adopt a standard web application threat model. The attacker is an arbitrary Internet user, in full control of their user agent, with the capability to craft HTTP requests to a web application fronted by a server-side cache. The attacker cannot control cache admission or replacement and cannot select which legitimate entry is evicted. Whether the transport is secure is irrelevant to our work.

\section{Methodology}
\label{sec:method}

\algrenewcommand\algorithmicfunction{\textbf{thread}}
\algrenewcommand\algorithmicprocedure{\textbf{do parallel}}
\begin{algorithm}
    \footnotesize
	\caption{Pseudo-code for the high-level \attack flow.}
    \label{alg:method}
	\textbf{Input} $target$: Victim cache-fronted website.

	\begin{algorithmic}[1]
    \State $path, keySpec \gets findObjectWithImpreciseKey(target)$
    \\
    \State \# Procedure 1: Fill the cache with redundant objects
    \State $busterList \gets []$
    \Repeat
        \State $busterRequest \gets$ $generateCacheBuster(path, keySpec)$
        \State $sendGET(busterRequest)$
        \State $busterList.append(busterRequest)$
    \Until{$isCacheFull() = True$}
    \\
    \State \# Procedure 2: Refresh the TTL of cached objects
    \Loop
        \ForAll{$busterRequest\in busterList$}
            \State $response \gets sendHEAD(busterRequest)$
            \If{$response.cacheStatus = Miss$}
                \State $sendGET(busterRequest)$
            \EndIf    
        \EndFor
    \EndLoop

    \end{algorithmic}
\end{algorithm}

We now describe how to achieve \attack in concrete steps. Conceptually, \attack is straightforward, but there are important considerations and optimization opportunities hidden in the details. We first provide a simplified view into the core methodology for brevity, and later discuss those considerations.

\subsection{Overview}
\attack follows the steps below, also illustrated in Algorithm~\ref{alg:method} with specific lines referenced in the text.
\begin{enumerate}
    \item Pick a target victim website fronted by a web cache. Prior knowledge of the cache capacity, cache eviction algorithm, or traffic patterns to the website is \textbf{not} necessary. However, if this information is available, it can be used to optimize the attack.
    \item Identify a cacheable object specified by an imprecise cache key (line 1). This is the object that the attack will fill the cache with, and large files are better.
    \item Abuse the imprecise cache key definition to generate a unique cache buster; i.e., a request for the cacheable object with the appropriate keyed element modified to cause a cache miss, resulting in a fresh copy of the object being fetched from the origin and stored under a new cache key. Repeat this process until the cache is approximately filled to capacity with redundant objects. Also, keep a list of all generated cache busters (lines 3-9).
    
    At this stage, the origin is suffering the full consequences of the attack due to cache degradation, resulting in repeated cache misses for legitimate traffic. The number of entries in the list of cache busters we maintain matches the number of objects needed to fill the cache.
    \item To maintain the attack and prevent legitimate traffic from eventually recovering the cache locality for popular objects, cycle through the list of cache busters used for filling the cache in the previous step, and probe them with HEAD requests (lines 11-19). The HEAD request refreshes the cache time-to-live (TTL) of the bogus object, preventing it from getting evicted.
    \item If the response for any HEAD request indicates a cache miss, that implies that the corresponding bogus object was already evicted. Therefore, restore it to the cache by sending a new GET request with the same cache buster (lines 15-16). 
\end{enumerate}

In Algorithm~\ref{alg:method}, the routines \textit{sendGET} and \textit{sendHEAD} are self-explanatory. We discuss the undefined routines \textit{findObjectWithImpreciseKey} and \textit{isCacheFull} later in the following subsections.

\subsection{Cost Considerations}

As with any volumetric attack, \attack is only meaningful when the damage inflicted on the origin server outweighs the cost incurred by the attacker. We define the metric \textit{Traffic Amplification Ratio (\metric)} to measure the effectiveness of the attack based on the disruption of the cache hit rate. \metric is calculated by measuring the increased traffic that reaches the origin server due to cache misses resulting from the attack, divided by the attacker's bandwidth expense. A higher value indicates greater disruption relative to the attacker's resource investment.

\[
\metric = \frac{T_{\text{attack}} - T_{\text{normal}}}{T_{\text{\attack}}}
\]

where:

\begin{itemize}
    \item \( T_{\text{normal}} \): Volume of traffic reaching the origin server due to cache misses under normal conditions.
    \item \( T_{\text{attack}} \): Volume of traffic reaching the origin server due to cache misses during the attack.
    \item \( T_{\text{\attack}} \): Traffic generated by the attacker.
\end{itemize}

The target cache's capacity and the size of the redundantly stored object directly factor into the cost, as they influence the number of requests necessary to fill the cache and maintain the attack. Consequently, the use of HEAD requests during attack maintenance is intentional. Since responses for HEAD requests do not contain the body payload, this greatly reduces client-side bandwidth and the attack cost. \metric measures attacker efficiency rather than absolute service degradation. Consequently, a configuration can retain a comparatively higher hit rate yet produce a larger \metric when maintaining the attack requires substantially less attacker traffic.

Beyond these general considerations, cache implementation behavior may further reduce client-side bandwidth. We identified two such cases in our experiments.

First, servers may be configured to perform opportunistic caching by upgrading a client's HEAD requests to GETs before forwarding them to the origin and storing the resulting object without transferring it back to the client. All GET requests in our methodology can be replaced with HEADs and lines 15-17 in Algorithm~\ref{alg:method} eliminated, resulting in the same disruption, but without making the attacker incur the cost of receiving the response payload. We observed this behavior with the default configurations of \nginx and Varnish.

Second, servers may allow for early termination of connections while still caching the response, allowing the attacker to issue GET requests but not read the response, resulting in a similar bandwidth optimization. In our experiments, Squid was the sole server that required the client to consume the entire body before caching an object, while all other caches allowed early termination.

Also note that while the legitimate traffic received at the website will compete with the attacker's probes for cache space and result in some bogus objects getting evicted until they can be restored via the TTL refresh loop (Steps 4--5 in the overview, lines 11--19 in Algorithm~\ref{alg:method}), that does not imply better traffic offload for the origin. Since the cache is still at storage capacity, even this temporary caching of genuinely popular content will result in cache thrashing and fail to remediate the cache degradation. We demonstrate this effect later in our evaluation, in Section~\ref{sec:evaluation}.

\subsection{Finding Objects With Imprecise Cache Keys}
\label{sec:method:finding-objects}

Identifying a cacheable object with an imprecise cache key definition, ideally a large one, hosted on the target website, is an essential part of \attack. Thankfully, this process can be automated. Our approach for identifying such exploitable objects leverages the methodology previously presented by Golinelli et al. in their work that describes how to detect hidden web caches, and extends it for our needs~\cite{hwcd}.

Specifically, we implement a recursive web crawler that is seeded by the target website's domain. The crawler issues GET requests for the resources linked from the crawled pages, and checks the response for indications of caching. These include response header heuristics compiled from the work of Mirheidari et al.~\cite{mirheidari2022web}, supplemented by us to include the headers introduced in RFC 9211~\cite{rfc9211}.

Once the crawl is complete, yielding a list of cacheable objects, we sort these by object size in descending order, and work through the list to automatically reverse engineer their cache key definitions. This process involves crafting a GET request for that object, systematically modifying the frequently keyed HTTP elements in isolation, sending the mutated request, and checking the response for cache status indicators using the aforementioned header heuristics. In particular, we test the following, in the given order:

\begin{enumerate}
    \item Modify the query string.
    \item Include and modify the headers drawn from keyed fields observed in default configurations of popular cache technologies (i.e., Origin, User-Agent, X-Forwarded-Host, X-Method-Override, X-Forwarded-Scheme, Accept-Language, Accept-Encoding, Accept, X-Forwarded-For).
    \item Edit the values of headers specified with the \texttt{Vary} header.
    \item Include \texttt{Cache-Control: no-cache}.
    \item Add arbitrary cookies.
\end{enumerate}

For each mutation, we compare the response representation with the baseline and inspect the cache-status indicators. We stop after finding one object for which an equivalent response can be stored under a distinct key because this establishes the prerequisite condition for the attack.

This is depicted on line 1 of Algorithm~\ref{alg:method} with the routine \textit{findObjectWithImpreciseKey}, which returns both the path for the exploitable object and the imprecise cache key definition, allowing us to generate an arbitrary number of cache busters by varying the appropriate keyed HTTP element.

\subsection{Detecting When the Cache Is Full}
Finally, an approximation of when the cache is full and the attack is at its peak is an important signal for the attacker. On one hand, over-approximating the number of redundant objects to cache is costly, as generating new cache busters and sending GET requests for them incur a higher cost than the TTL maintenance loop of sending HEAD requests, providing no real benefit in the process. On the other hand, under-approximating this number means that the attacker switches to the maintenance loop too early, leaving space in the cache for genuinely popular objects, and hence under-utilizing \attack. We propose two methods to tackle this problem.

First, if the cache capacity can be estimated, dividing it by the known stored object size approximates the number of entries required to occupy that capacity. This information may not be known for proprietary deployments. However, note that many production systems rely on pre-packaged software with baked-in defaults (e.g., caching proxies deployed via containers in public registries, managed cache components provided by hyperscalers like AWS, Azure, GCP), making informed estimations practical if contextual clues for the deployment setup exist. We use a similar methodology in our evaluation, in Section~\ref{sec:evaluation}, to collect information on commonly seen cache sizes.

Second, when cache capacity estimation is not feasible, attackers can monitor cache hit rate. Specifically, one can select a highly popular cacheable object (e.g., an image banner on the homepage), periodically issue requests for that object while the attack is in progress, and monitor the cache hits and misses. Observing a cache miss for an object served from the homepage is unlikely and could alone indicate that the attack has escalated to high severity. The attacker can then decide whether to cache more objects or switch to TTL maintenance depending on the cache hit/miss rate calculated over these monitoring probes.

\section{Evaluation}
\label{sec:evaluation}

We now evaluate how \attack performs in practice through a set of detailed experiments, and start answering the research questions we laid out in Section~\ref{sec:research}.

Testing \attack on production websites is not ethically feasible. DoS attacks are explicitly prohibited on major bug bounty platforms, and even throttled volumetric tests can impose performance and cost consequences on website operators.

We therefore evaluate the attack in a controlled laboratory environment, attacking our own infrastructure. This is aligned with all the cache pollution works we cited in Section~\ref{sec:background}, and is the overall standard approach in DoS research. To ensure our evaluation methodology is realistic, we first perform non-disruptive Internet measurements and use the observed object sizes and configured cache capacities to select parameters for our controlled experiments.

\subsection{Imprecise Cache Keys and Object Size In The Wild}
\label{sec:large-scale-file-size}

Identifying cacheable objects with imprecise cache keys is the prerequisite for \attack. Therefore, we first perform a large-scale Internet experiment to determine whether imprecise cache keys indeed exist in popular websites. While doing so, we also measure the sizes of these cache-bustable objects, as this is a critical parameter that influences the viability of \attack that we must evaluate later.

We conduct this measurement on the Tranco top 10k domain list generated on April 7, 2025~\cite{tranco}.\footnote{Available at \url{https://tranco-list.eu/list/LJL44}.} We follow the methodology in Section~\ref{sec:method:finding-objects}: we crawl each website, identify cacheable objects, and automatically reverse engineer their cache keys. We limit the crawl to a maximum of 10 pages on each of at most 10 unique subdomains per website to avoid generating unreasonable traffic volumes. During cache-key inference, we process objects by size in descending order and stop when we find the largest object with an imprecise cache key. Our objective is not to perform a comprehensive measurement study detailing all exploitable objects, but to demonstrate the existence of at least one such object on a website, therefore minimizing the experiment's burden on production systems.

Out of the 10k websites in our dataset, we were able to crawl 5437, while the rest did not respond to our HTTP requests because of infrastructure issues or possible bot-detection defenses. Among the websites successfully tested, 4000 contained cacheable objects, and \textbf{3600} of these contained at least one object with an imprecise cache key. We did not detect the condition for any tested object on the remaining 400 websites.
The five-number summary for the affected objects is as follows: $Min=2\,\textrm{bytes}, Q1=0.12\,\textrm{MB}, Median=0.37\,\textrm{MB}, Q3=1.02\,\textrm{MB}, Max=187.54\,\textrm{MB}$.

We note that these numbers are lower bounds due to the limits we imposed on the crawl for ethical considerations (e.g., we did not make attempts to bypass bot detection techniques, which are commonly utilized for large file downloads). Case in point, our experiment did not catch a 6 GB operating system image hosted on one of the tested sites, which we later manually confirmed to have an imprecise cache key.

To empirically assess this lower bound, we randomly sampled 100 websites from the subset with affected objects smaller than the median size and manually searched each site for larger objects exhibiting the same condition. 16 sites redirected to other domains and were discarded. On 72 sites, we found larger affected objects; on 12 sites, we did not. The five-number summary of size ratios between the largest manually discovered affected object and the automatically detected one is: $Min=2.96, Q1=43.84, Median=148.53, Q3=1605.43, Max=59507.35$. This indicates that in many cases, significantly larger cacheable objects with imprecise cache keys exist on the same website.

This confirms that imprecise cache keys are indeed very common, and they can be automatically detected, answering our first research question affirmatively (i.e., Section~\ref{sec:questions}, Q1).
As for the object size distribution, we confirm that more than 25\% of websites have exploitable objects larger than 1 MB, extending into the 100 MB range. However, the median is fairly small at 0.37 MB. We leverage the insights from this measurement in the rest of our evaluation, and test how these object sizes influence the success of \attack.

\subsection{Cache Capacity In The Wild}

To establish realistic cache capacity values to use in our experiments, we selected five popular open-source caching proxy technologies (i.e., Apache Traffic Server (ATS), HAProxy, \nginx, Squid, and Varnish), and searched GitHub for projects that use them via the GitHub API. We then analyzed the cache configurations these projects are bundled with.

This process yielded 127 projects with valid cache configurations; a sample list is available in Appendix~\ref{app:popular-repos}. Table~\ref{tab:cache-storage} summarizes the results, showing that the median cache capacity does not exceed 1 GB for any proxy in this dataset. Most observed configurations were below 50 GB, with a small number of ATS outliers exceeding 100 GB. We designed the rest of the experiments in this section in light of these findings.

\begin{table*}[t]
    \caption{Cache capacities configured in the analyzed GitHub projects. The \# column gives the number of configurations. Min and Max are the smallest and largest configured values observed in this dataset.}
    \label{tab:cache-storage}
    \centering

    \begin{tabular}{lrrrrrr}

    \toprule

    Proxy   & \#  & Min (MB)   & Q1 (MB)   & Median (MB)   & Q3 (MB)   & Max (MB)   \\

    \midrule
    ATS     & 23       & 144      & 256     & 256         & 10240   & 1048576  \\
    HAProxy & 34       & 4        & 20      & 164         & 512     & 4095     \\
    \nginx   & 29       & 73       & 256     & 1024        & 1024    & 51200    \\
    Squid   & 21       & 16       & 100     & 512         & 5000    & 51200    \\
    Varnish & 20       & 1        & 256     & 256         & 1280    & 32768    \\
    \bottomrule
    \end{tabular}
\end{table*}

\subsection{\attack With Common Defaults}

We perform our first \attack experiment with the same five caching proxies. This experiment tests whether the redundant-entry mechanism produces cache degradation across implementations under one uniform configuration informed by the preceding measurements.

We configure all proxies to cache every response. We work with objects of size 1 MB, and we set the cache capacity to 50 GB, both informed by our observations in the previous experiment. The only exception is HAProxy, for which the maximum cache capacity is 4095 MB. Other configuration parameters are left at their default settings. In particular, all caches employ the Least Recently Used (LRU) eviction algorithm, and they define their cache keys to include the query string, allowing us to perform cache busting by modifying its value.

To simulate the routine traffic a website receives, we use Web Polygraph, a benchmarking tool for caching proxies and other web intermediaries~\cite{rousskov2004high}. Web Polygraph provides a client component that generates traffic and a server component that responds to it. We place the tested cache between them and measure the cache hit rate.

Web Polygraph provides predefined workloads simulating common traffic patterns. We use the workload \texttt{simple.pg}.\footnote{\url{https://github.com/albertok/web-polygraph/blob/master/workloads/simple.pg}} This workload represents a mix of repeated and non-repeated requests with a configurable request rate, which directly affects the cache hit rate during the experiment. By default, the recurrence rate is set to 55\%, meaning that Web Polygraph requests the same resources 55\% of the time. Since we set all responses to be cacheable, we anticipate a consistent hit ratio of around 55\%. We configure Web Polygraph to issue 100 requests per second.

We test each cache in isolation by deploying it as a reverse proxy in our lab environment, in front of our origin server that serves static files of varying sizes. The attacker's requests are routed to this origin, while the simulated routine traffic is directed to Web Polygraph's server. The two routes share the same cache. As a result, the impact of \attack manifests in Web Polygraph's benchmarks.

For each experiment, we first run Web Polygraph for 10 minutes without performing any attack to warm the cache. We then run \attack for the next 30 minutes. During both phases, we record the cache hit rate. Compared to the subsequent experiments, we use a longer attack duration here to clearly illustrate the attack's warm-up and maintenance phases.

\begin{table}[t]
    \caption{Cache hit rate (HR) before and during \attack. Values are averages over the 10-minute phases.}
    \label{tab:dos}
    \centering

    \begin{tabular}{lrr}

    \toprule

    {Cache} & \multicolumn{1}{c}{Before HR (\%)} & \multicolumn{1}{c}{During HR (\%)} \\

    \midrule

    ATS     & 55.1 & 30.0 \\
    HAProxy & 55.1 & 0.5  \\
    \nginx  & 55.1 & 6.9  \\
    Squid   & 55.0 & 32.9 \\
    Varnish & 55.0 & 7.2  \\

    \bottomrule
    \end{tabular}
\end{table}

\begin{figure}[t]
    \centering
    \includegraphics[width=1.0\linewidth]{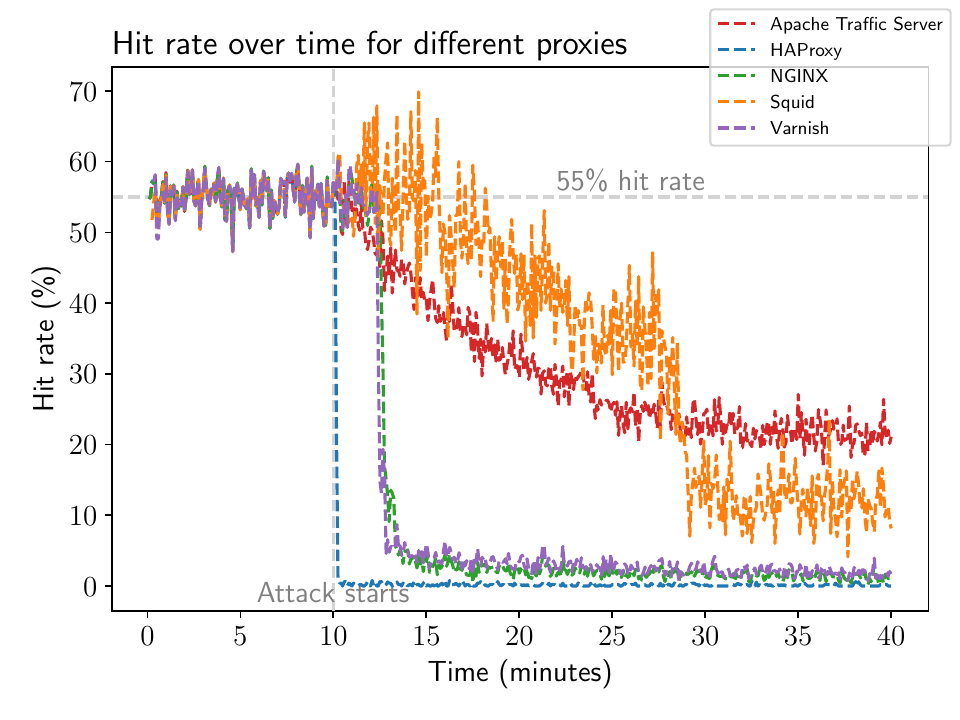}
    \caption{Cache hit rate before and during \attack with different caching technologies. The cache size is 50 GB, while the object size is 1 MB.}
    \label{fig:dos-proxies}
    \Description{A line graph showing the cache hit rate before and during \attack with different caching technologies. The x-axis represents time, while the y-axis represents the cache hit rate. Each line corresponds to a different caching technology, and they all show a significant drop in cache hit rate during the attack phase.}
\end{figure}

Table~\ref{tab:dos} presents the average cache hit rate for each proxy before and during \attack. The results show that \attack significantly degrades performance in all cases. The cache hit rate drops further with the increase in the origin server's traffic load. Some caches retain a somewhat higher residual hit rate under our setup, but we do not interpret this as inherent resilience. Rather, these differences likely reflect implementation-specific behavior under the single attack configuration we applied uniformly across all caches. A more sophisticated, cache-specific attack strategy could plausibly shift these relative results further. Figure~\ref{fig:dos-proxies} visually depicts the cache hit rate drop as the attack progresses.

The results so far are promising, demonstrating that \attack renders the cache ineffective under widely used default configurations.

\subsection{\attack With Varying Parameters}

We next evaluate how cache capacity and cached object size influence the outcome. Having reproduced the mechanism across five cache technology implementations, we use \nginx for the remaining controlled experiments.

In these experiments, we also measure bandwidth and calculate the Traffic Amplification Ratio (\metric) as the parameters vary, to realistically validate that \attack is a practical DoS attack. During maintenance, we throttle the attack traffic so that every redundant entry is probed approximately once every 60 seconds, rather than sending requests as quickly as possible. This interval provides a consistent basis for comparing attacker efficiency; different TTLs, admission policies, and legitimate traffic would change the maintenance rate required in a deployment.

\subsubsection{Cache Capacity}
\label{sec:cache-storage-size}

\begin{table}[t]
    \caption{\attack cost with an object size of 1 MB and varying cache capacity.}
    \label{tab:cache-sizes}
    \centering

    \begin{tabular}{@{}rrrrr@{}}
    \toprule
    Cache Capacity & \multicolumn{1}{c}{$T_{\text{normal}}$} & \multicolumn{1}{c}{$T_{\text{attack}}$} & \multicolumn{1}{c}{$T_{\text{\attack}}$} & \multicolumn{1}{c}{\metric} \\

    \multicolumn{1}{c}{(GB)} & \multicolumn{1}{c}{(MB)} & \multicolumn{1}{c}{(MB)} & \multicolumn{1}{c}{(MB)} & \multicolumn{1}{c}{} \\

    \midrule

    100    & 351.37     & 678.25     & 153.77     & 2.13 \\
    50     & 348.42     & 684.02     & 133.55     & 2.51 \\
    10     & 349.98     & 707.79     & 36.05      & 9.92 \\
    1      & 348.76     & 618.35     & 3.64       & 74.05 \\

    \bottomrule
    \end{tabular}
\end{table}

\begin{figure}[t]
    \centering
    \includegraphics[width=1.0\linewidth]{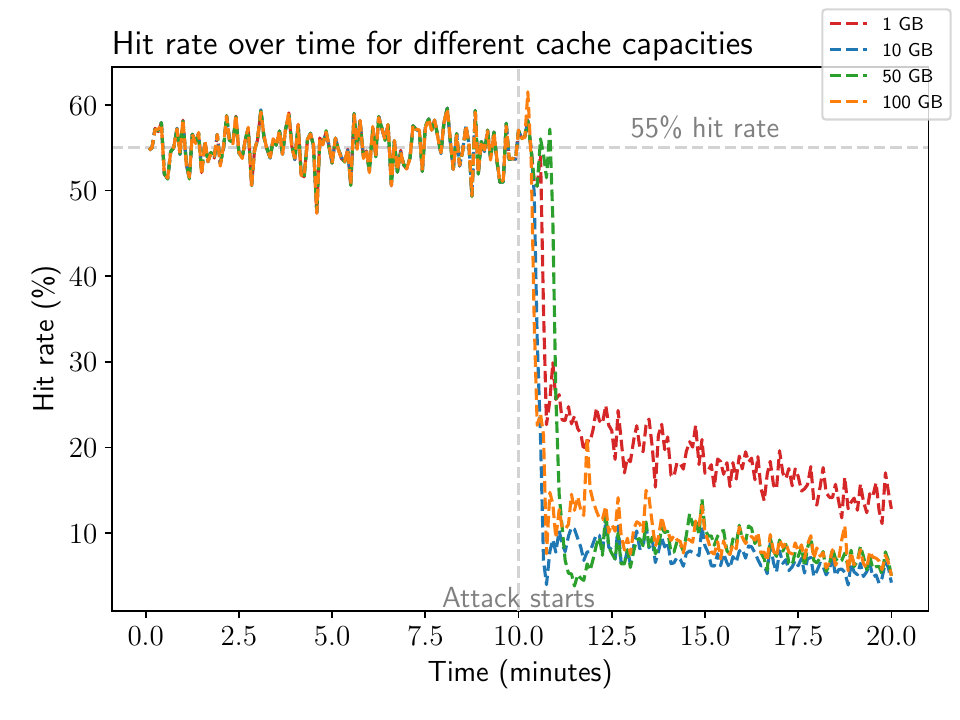}
    \caption{Cache hit rate before and during \attack with \nginx for various cache capacities and an object of 1 MB.}
    \label{fig:cache-storage-size}
    \Description{A line graph showing the cache hit rate before and during \attack with \nginx for various cache capacities and an object of 1 MB. The x-axis represents time, while the y-axis represents the cache hit rate. Each line corresponds to a different cache capacity, and they all show a significant drop in cache hit rate during the attack phase.}
\end{figure}

We run four tests with \nginx configured with a cache storage capacity ranging from 1 GB to 100 GB, informed by our previous measurement experiment. We also repeated the attack using several objects of different sizes in our exploratory studies, ranging from 100 KB to 500 MB; here, we present the results with a 1 MB file.

Figure~\ref{fig:cache-storage-size} shows the cache hit rate before and during the attack, once again demonstrating that \attack rapidly degrades performance in the same way, regardless of the cache capacity. However, we also see in Table~\ref{tab:cache-sizes} that the cache capacity has a notable impact on the attacker's cost. Larger caches require more objects to fill, which in turn translates to higher bandwidth. We stress that this is a conservative case. As we show in the next experiment, the attacker could employ larger objects to perform the attack efficiently on a 100 GB cache. Recall that the attacker does not need to download the full body of each response, as explained in Section~\ref{sec:method}, which is why the bandwidth values are significantly lower than the cache capacity.

\subsubsection{Object Size}
\label{sec:file-size}

We now configure \nginx with a cache capacity of 50 GB and perform \attack using four different object sizes, ranging from 100 KB to 500 MB.

\begin{figure}[t]
    \centering
    \includegraphics[width=1.0\linewidth]{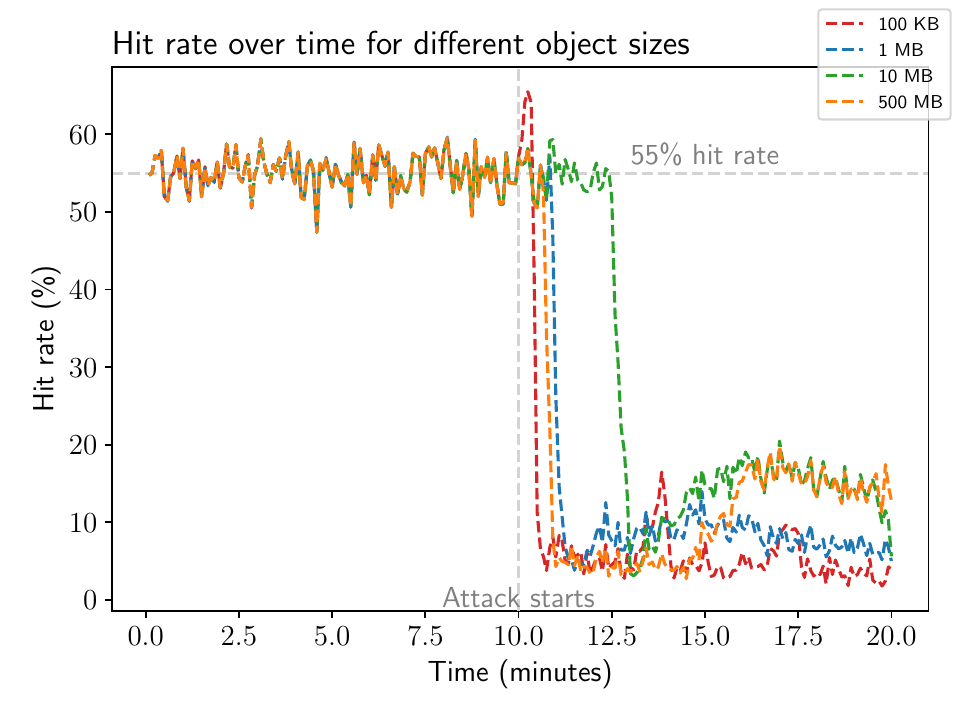}
    \caption{Cache hit rate before and during \attack with \nginx for different object sizes and cache capacity of 50 GB.}
    \label{fig:object-size}
    \Description{A line graph showing the cache hit rate before and during \attack with \nginx for different object sizes and cache capacity of 50 GB. The x-axis represents time, while the y-axis represents the cache hit rate. Each line corresponds to a different object size, and they all show a significant drop in cache hit rate during the attack phase.}
\end{figure}

Figure~\ref{fig:object-size} shows the resulting cache hit rate, demonstrating that \attack works regardless of object size. Table~\ref{tab:object-sizes} confirms our intuition that larger objects yield far better \metric values, but even with a 1 MB object, \attack remains viable.

\begin{table}[t]
    \caption{\attack cost with a cache capacity of 50 GB and varying object sizes.}
    \label{tab:object-sizes}
    \centering

    \begin{tabular}{@{}rrrrr@{}}
    \toprule
    Object Size & \multicolumn{1}{c}{$T_{\text{normal}}$} & \multicolumn{1}{c}{$T_{\text{attack}}$} & \multicolumn{1}{c}{$T_{\text{\attack}}$} & \multicolumn{1}{c}{\metric} \\

    \multicolumn{1}{c}{(MB)} & \multicolumn{1}{c}{(MB)} & \multicolumn{1}{c}{(MB)} & \multicolumn{1}{c}{(MB)} & \multicolumn{1}{c}{} \\

    \midrule

    500    & 357.82     & 676.78     & 15.31     & 20.84 \\
    10     & 351.92     & 583.28     & 7.76      & 29.83 \\
    1      & 348.42     & 684.02     & 133.55    & 2.51 \\
    0.1    & 350.65     & 571.81     & 391.43    & 0.57 \\

    \bottomrule
    \end{tabular}
\end{table}

\subsection{Eviction Algorithms}

LRU is the most common, and often the only supported eviction algorithm, with many web cache technologies. Yet, some of the works we cited in Section~\ref{sec:background} specifically explore the use of other caching strategies in their defenses against cache pollution (e.g.,~\cite{icn10}). Therefore, we also evaluate the impact of different eviction algorithms on the effectiveness of \attack.

For this particular experiment, we use Squid instead of \nginx, since only Squid supports the variety of eviction algorithms we would like to test. We set the cache capacity to 50 GB, and the cached object size to 1 MB. We then run tests for each eviction algorithm that Squid supports: LRU, heap LRU, heap GDSF, and heap LFUDA.

Table~\ref{tab:eviction-algorithms} presents the results of this experiment. In short, we saw no difference of note between these different eviction algorithms, and we conclude that \attack is robust against tweaks to the caching strategy. This confirms our previous claim that cache eviction details are irrelevant for \attack's abuse mechanism of filling a cache to capacity.

\begin{table}[t]
    \caption{\attack impact on Squid with different eviction algorithms.}
    \label{tab:eviction-algorithms}
    \centering

    \begin{tabular}{lrr}

    \toprule

    {Eviction Algorithm} & \multicolumn{1}{c}{Before HR (\%)} & \multicolumn{1}{c}{During HR (\%)} \\

    \midrule

    Heap LRU   & 55.3 & 36.7 \\
    LRU        & 55.3 & 34.2 \\
    Heap GDSF  & 55.3 & 35.2 \\
    Heap LFUDA & 55.3 & 35.2 \\

    \bottomrule
    \end{tabular}
\end{table}

\subsection{\attack-facilitated Cache Poisoning}
\label{sec:cache-poisoning}

We next evaluate \attack as an eviction-pressure mechanism that may compose with an independent cache-poisoning vulnerability. This experiment is a proof of composition against a deliberately vulnerable laboratory application.

We add to our setup an origin application containing a reflected XSS vulnerability whose response is cacheable and poisonable. We use a 1 MB object and vary cache capacity. We restrict this experiment to smaller caches because the attack did not succeed within a practical experimental duration for larger configurations. Attackers with more resources or larger objects could overcome this limitation and poison larger caches.

We start the test in a state where a response for the vulnerable endpoint is already cached, as it would be in a real-world scenario.
The vulnerable endpoint regularly receives requests from simulated legitimate clients. The attacker generates redundant entries while repeatedly attempting to exercise the independent poisoning vulnerability. The attacker cannot select which object is evicted. If the target becomes absent, the attacker has an opportunity to store the malicious response; a legitimate request may instead reinsert the benign response, requiring the attacker to maintain eviction pressure and wait for another opportunity.

We let this scenario play out with simulated traffic representing arbitrary clients. The attacker performs \attack as usual, while simultaneously trying to poison the vulnerable endpoint, competing with regular traffic. We experiment with various routine request rates for the endpoint, and we perform 5 trials for each of these experiments to measure the average time for a successful poisoning attack.

Table~\ref{tab:cache-poisoning} shows that success depends on both cache capacity and legitimate request rate. Several 4 GB configurations did not succeed within the 15-minute timeout, and the mechanism did not select the target for eviction. In this deliberately vulnerable laboratory application, redundant-entry flooding increased the opportunities for a previously cached endpoint to become absent, facilitating exploitation of previously immaterial vulnerabilities.

\begin{table}[t]
    \caption{Average time to successfully perform cache poisoning by leveraging \attack to purge a target object with different cache capacities and varying regular traffic rates. ``-'' means the attack did not succeed within a 15-minute timeout.}
    \label{tab:cache-poisoning}
    \centering

    \begin{tabular}{c@{\hspace{15pt}}r@{\hspace{15pt}}r@{\hspace{15pt}}r}

    \toprule

    \multicolumn{1}{c}{Request Rate (r/s)}  & \multicolumn{3}{c}{Average Time to Success (s)} \\

    \multicolumn{1}{c}{} & \multicolumn{1}{c}{1024 MB} & \multicolumn{1}{c}{2048 MB} & \multicolumn{1}{c}{4096 MB} \\

    \cmidrule(r){1-1} \cmidrule(r){2-2} \cmidrule(r){3-3} \cmidrule(r){4-4}

     1     & 32.0        & 49.0       & 15.8    \\
     10    & 30.2        & 50.0       & -       \\
     50    & 18.0        & 63.8       & -       \\
     100   & 92.6        & 108.4      & -       \\

    \bottomrule
    \end{tabular}
\end{table}

\subsection{Summary of Results and Limitations}
Our measurements over Tranco top 10k show that more than one third of these popular websites have cacheable objects with imprecise cache keys, confirming the premise for \attack.

The remaining experiments demonstrate that the core mechanism of \attack, that is, filling caches with duplicate objects, forcing the eviction of everything else, and leaving the cache in a perpetual state of cache thrashing, is viable with all tested caching proxies, regardless of the parameters such as cache capacity and exploitable object size. These parameters, however, do have practical impacts on whether \attack is viable as a DoS vector. Extreme values (very small files, very large caches) can become cost-prohibitive, resulting in low \metric values. This is an inherent limitation of the mechanism and methodology we propose in this research, and there are cases where \attack may not be an effective attack. However, as also evidenced by our study of commonly seen object sizes and cache capacity configurations in the wild, such extreme values for these parameters are not the norm, and there are many opportunities to launch viable \attack attacks in realistic web application deployment settings.

All in all, we conclude that the findings we presented here successfully address our research questions Q1, Q2, and Q3 from Section~\ref{sec:questions}, establishing \attack as a practical threat.

\section{Mitigations}
\label{sec:mitigations}

To answer our last research question Q4, we describe several mitigation strategies for \attack and test their effectiveness.

\subsection{Cache Deduplication}
\label{sec:cache-de-duplication}

An effective mitigation against \attack is deduplication. Cache deduplication may be implemented by computing checksums of cacheable responses as they are received, comparing them against the checksums of already cached objects, and only storing a new copy if the object is unique in storage. Otherwise, a data structure would be updated to associate the distinct cache keys with a single stored copy of the object.

While deduplication is a well-known technique designed for this exact problem, its application to web caches could be challenging. Both the checksum computation and the cache key tracking data structure management tasks can become prohibitively costly for a busy server, and worse, cause processing delays, increasing application response times.

We run a simple experiment to estimate this cost by checksumming files of various sizes and measuring the time the operation takes. We test a cryptographic hash (i.e., SHA-512), and also a non-cryptographic checksum (xxHash, specifically the XXH3 variant~\cite{xxHash}) designed to be very fast to compare. We perform 1000 trials for every experiment, and present the average times elapsed.

Table~\ref{tab:deduplication} summarizes the results. This may appear negligible in isolation, but given that today's production platforms process millions of requests per second (e.g.,~\cite{reqs}), processing cycles and latency may quickly become unreasonable.

\begin{table}[t]
    \caption{Average CPU time over 1000 runs to compute file hashes and checksums.}
    \label{tab:deduplication}
    \centering

    \begin{tabular}{rrr}
    \toprule

    {}            & xxHash            & SHA-512           \\
    {File Size}   & Avg. CPU Time (s) & Avg. CPU Time (s) \\

    \midrule

    500 MB  &   0.24258   &  0.81138  \\
    10 MB   &   0.00548   &  0.02170  \\
    5 MB    &   0.00288   &  0.01296  \\
    1 MB    &   0.0007    &  0.00304  \\
    500 KB  &   0.0004    &  0.00161  \\
    100 KB  &   0.00013   &  0.00042  \\

    \bottomrule
    \end{tabular}
\end{table}

\subsection{Anomaly Detection}

Anomaly detection is the main defense approach adopted by nearly all cache pollution defense works we referenced in Section~\ref{sec:background}. Most relevant for our context, the works by Gao et al. and Deng et al. that present cache pollution attacks and defenses on forward web caches propose a detection scheme based on various request and traffic features~\cite{gao2006internet,deng2008pollution}. These works show that their approach is effective, but it also comes with processing overhead and false positives. Moreover, it can take up to 10 hours to catch anomalies. These downsides are not unique to the cited works and are expected of any anomaly detection approach to security.

\attack can also be detected in this manner, likely more easily than other cache pollution attacks, due to its noisier nature. If the shortcomings of anomaly detection are acceptable, this could be a viable solution. However, we also expect the costs to be material for a real-life application, and as discussed in the cited works, IP blocking capabilities themselves may be weaponized by an attacker to block NAT gateways. All in all, we do not believe anomaly detection should be the preferred approach if a more effective solution can be found.

\subsection{Rate Limits}
\label{sec:rate-limits}

\begin{table}[t]
    \caption{\attack hit rates with different rate limits, a cache of 50 GB and an object of 1 MB.}
    \label{tab:rate-limits}
    \centering

    \begin{tabular}{crr}

    \toprule

    {Rate Limit (req/s)} & \multicolumn{1}{c}{Before HR (\%)} & \multicolumn{1}{c}{During HR (\%)} \\

    \midrule

     50 & 55.1 & 54.9 \\
     10 & 54.6 & 54.6 \\
     5  & 38.4 & 37.8 \\
     1  & 11.7 & 10.1 \\

    \bottomrule
    \end{tabular}
\end{table}

\attack generates a large volume of sustained traffic over a long time for the TTL maintenance loop. Therefore, we explore rate limiting separately from more advanced anomaly detection as a less costly option.
We conduct an experiment with the same evaluation setup to test this approach. We use \nginx, set the cache capacity to 50 GB, and employ a 1 MB file as the cached object. We configure Web Polygraph to issue 10 requests per second from 10 different IP addresses, for a total of 100 requests per second. We run Web Polygraph for 10 minutes without performing any attack, and then run it together with \attack for another 10 minutes. In both phases, we measure the cache hit rate. We repeat this experiment while applying rate limits ranging from 1 to 50 requests per second, calibrating attack intensity to each limit. If the attacker issues requests faster than the configured limit, the rate limiter will begin dropping excess requests. Those drops will discard many of the attacker's requests and prevent the sustained traffic required to fill the cache. Conversely, an attacker that reduces its rate to be at or below the measured limit can avoid being dropped and, although the attack proceeds more slowly, given sufficient time, it can still gradually populate the cache. In our experiments, we calibrate the attack rate to the applied limit. In the real world, an attacker could easily spot and measure the rate limit and adjust accordingly.

Table~\ref{tab:rate-limits} shows the cache hit rate before and during the attack. The results indicate that rate limiting is effective in reducing the attack's impact. That said, rate limits below 10 requests per second, while being effective in reducing the impact of the attack, naturally also impact regular traffic, causing a drop in the cache hit rate before the attack begins. This once again demonstrates that rate limits cannot be applied generally to all traffic without hurting normal traffic flows, and just like anomaly detection, false positives can be disruptive, or they can even be weaponized by an attacker to get legitimate clients blocked.

\subsection{Stricter Cache-Key Design}

Abuse of imprecise cache keys is a necessary precondition for \attack. That also makes eliminating imprecise cache keys the optimal mitigation. In light of all the conceptualization, discussion, and experiments we presented in this work, we argue that this is the preferred approach.

The main drawback of this approach is that it requires website operators to thoroughly analyze their cache configurations and apply the necessary corrections. At the same time, the same imprecise cache detection methodology we presented in Section~\ref{sec:method:finding-objects} can be repurposed to run deep crawls on websites and identify all problem cache keys, automating the bulk of the work. Revising the cache key still requires manual work informed by contextual knowledge of the surrounding application, but we argue that this cost is preferable over the challenges of sifting through false positives or suffering a sustained performance overhead.

Case in point, we developed a penetration testing tool that automates the systematic detection of imprecise cache key configurations on websites. The tool functions as a web crawler that recursively explores a target site, collecting and testing all accessible resources. For each resource, it issues a series of HTTP requests, each differing from the baseline request by a random modification to one element; specifically, the query string, selected headers, or cookies. The tool then applies response header heuristics derived from the work of Mirheidari et al.~\cite{mirheidari2022web} to determine whether a given modification caused the request to bypass the cache. If a modified request results in a cache miss while the corresponding response remains identical to the unmodified version, the tool infers that the cache key configuration is imprecise. The tool provides recommendations to the user on how to patch its findings using natural language. The tool is based on the methodology discussed in this paper and released as open source for researchers and practitioners to use freely in their own environments.

All in all, we assert that the consequences of imprecise cache keys should be communicated to website operators and cache administrators, and crafting strictly precise cache key definitions should be considered a best practice.

\section{Discussion}
\label{sec:discussion}

\subsection{Novelty}

We now reiterate the differentiating factors between \attack and prior cache pollution strategies. Traditional attacks rely on estimates or knowledge of cache access patterns, object popularity distributions, or replacement behavior. Our results show that \attack operates without such information, as it is solely based on identifying imprecise cache keys: a process that, as demonstrated in Section~\ref{sec:method} and Section~\ref{sec:large-scale-file-size}, can be trivially automated. This simplicity makes the attack both easier to deploy and more general in scope, removing the need for any target-specific reconnaissance or fine-tuned strategies. \attack actively generates arbitrary cache entries using cache-busting techniques, rather than relying on pre-existing unpopular resources, removing any constraints imposed by the amount and size of the existing content available on the victim site. \attack is in fact only limited by environmental factors, such as available bandwidth, server-side rate limiting, cache capacity, and the cache expiration policy.

Furthermore, leveraging lightweight HEAD requests instead of full object downloads enables sustained disruption with minimal effort and very low bandwidth and computational cost, making it accessible even to resource-limited adversaries. Finally, the consequences of \attack extend beyond mere cache performance degradation. Its ability to force eviction of cached objects enables new avenues for exploitation, such as cache poisoning or cache-based side channels, by removing the usual precondition that the targeted resource must not already be cached. We discussed how, for instance, an attacker can leverage \attack to create a window of opportunity for turning a reflected XSS vulnerability into a stored one through cache poisoning.
Overall, we argue that the results presented in this paper highlight that \attack is not a minor variation of existing cache attacks, but a fundamentally different threat model.

\subsection{What About Content Delivery Networks?}

In this paper, we focused our discussion on server-side caches, and in particular, for our evaluation, stand-alone caching proxies. This scope is not unrealistic; such caching reverse proxies deployed in multiple layers are ubiquitous and essential components of scalable web and cloud architectures. That said, Content Delivery Networks (CDNs) that operate massively distributed Internet overlays of caching reverse proxies also provide immense amounts of caching capacity to the web.

We cannot perform experiments with CDNs due to the aforementioned ethics considerations. Even an ineffective attempt would impose a cost on the infrastructure operators, and volumetric attack testing is explicitly prohibited in the Acceptable Use Policies of all major CDNs we checked.

Even without testing, we assume that \attack would not work on CDNs. While there are no public records of CDN cache sizes, a quick Internet search reveals many speculative accounts of CDNs deploying caches with capacities that are orders of magnitude larger than common stand-alone cache configurations. As evidenced by our evaluations, \attack does not yield a viable \metric value at such extreme values. Therefore, we operate under the assumption that \attack is not relevant for CDNs, and this is a fundamental limitation of our work.

To confirm this assumption and to notify them of our work, we contacted Akamai, AWS CloudFront, CDN77, Cloudflare, Fastly, Google Cloud CDN, KeyCDN, Bunny CDN, and OVHCloud. We shared our findings with them by providing a summary of this research, our proof-of-concept attack tools for their testing, and asking for their feedback.

Akamai, Cloudflare, Fastly, and Google acknowledged the research and reported that the attack would not be feasible at their scale. AWS stated that DoS attacks are outside its bug bounty scope (even though we did not seek a bounty) and provided no further comment. The remaining parties did not respond to contact attempts.

\section{Conclusion}
\label{sec:conclusion}

We presented \attack, a novel cache pollution attack that abuses imprecise cache keys to fill a web cache with duplicate objects, purging other cached content, and causing cache degradation. Our evaluation confirmed that imprecise cache keys are commonly found on popular websites, and \attack can effectively abuse them in several practical cache deployment setups. Moreover, we showed that \attack's capability to purge objects on command alone can be a valuable tool in the hands of adversaries. These findings address all four of our research questions, and confirm our hypothesis that \attack is a viable threat.

As we conclude, we reiterate that the crucial lesson learned from this research is that imprecise cache keys can have severe security implications. The resulting issues are best and most efficiently avoided by carefully designing precise cache keys as a general security best practice.

\balance

\bibliographystyle{plain}
\bibliography{main}

\appendix

\section{Ethical Considerations}
\label{sec:ethics}

All the attack experiments we present in this paper were conducted in realistic but controlled lab environments, without affecting external systems.

We also designed the Tranco top 10k experiments to minimize the traffic load on the targeted websites and to eliminate all damage outcomes. Specifically, we limited each crawl to at most 10 subdomains for each website and at most 10 pages per subdomain. We also limited our traffic rate to 1 request per second to avoid overwhelming the servers. To minimize the overhead of testing the websites for imprecise cache keys, we stopped on detecting the first successfully cache-busted object.

Most importantly, we note the following additional ethical considerations. The issue we present is a consequence of imprecise cache key configuration rather than a vendor-specific implementation vulnerability; consequently, disclosure to caching-proxy software vendors was not appropriate or likely to be actionable. More broadly, \attack reflects a configuration weakness, not a vendor-specific vulnerability. Cache key design is generally not treated as a security practice, and many operators may not explicitly manage cache keys, especially in managed, outsourced, or inherited deployments. While we identified imprecise cache keys, we did not validate the surrounding deployment context, such as whether a CDN was in use or whether other mitigations would make exploitation impractical. Reporting imprecise cache keys in isolation without validating exploitability risked creating confusion and overstating the practical severity for individual sites. We also did not perform vulnerability detection on production sites that use caching proxies, which would have required active experiments capable of causing DoS. Conducting such testing would have been unethical and out of scope, even for sites that publish vulnerability disclosure policies, so we refrained from active probing. Therefore, we did not identify or report any specific vulnerable sites.

Finally, we did not issue targeted notifications or a separate public advisory because, to our knowledge, no CERT-like organization accepts advisories covering broad configuration-class issues of this scope, and targeted notification without deployment-specific validation could overstate risk rather than help remediation. Instead, we consider broad documentation of the issue together with the release of detection tooling in this paper to be the most responsible and viable approach for operators and researchers.

\section{Generative AI Usage}
\label{sec:generative-ai}

Part of the code used in our experiments was developed with the assistance of AI-powered tools. These tools were employed to suggest code completions, boilerplate structures, and implementation details during development. All generated code was manually and thoroughly reviewed, tested, and adapted by the authors to ensure correctness and alignment with the intended experimental design.

\section*{Sample GitHub Repositories}
\label{app:popular-repos}

\begin{table*}[t]
    \caption{Samples of the 10 most popular GitHub repositories where we identified a cache configuration for the 5 selected proxies.}
    \label{tab:app-cache-storage}
    \centering

    \begin{tabular}{llrr}

    \toprule

    Cache      & Repository Name                             & Stars   & Size (MB)              \\

    \midrule

    ATS &  apache/trafficserver              &  1860  & 256 \\
    ATS &  mustafaramadhan/kloxo             &  336   & 256 \\
    ATS &  sqawasmi/trafficserver-docker     &  19    & 256 \\
    ATS &  Li4n0/My-CTF-Challenges           &  16    & 256 \\
    ATS &  sunnyszy/lrb-prototype            &  9     & 1048576 \\
    ATS &  ShufanWangBGM/Reinforcement-[...] &  8     & 524288 \\
    ATS &  xyp-root/geektime-hands-on-[...]  &  3     & 1024 \\
    ATS &  JasonGiedymin/ats-docker          &  2     & 256 \\
    ATS &  mingzym/zym4T                     &  2     & 144 \\
    ATS &  bryancall/benchmark               &  1     & 102400 \\

    HAProxy    & haproxy/haproxy                             & 5508    & 200.0              \\
    HAProxy    & haproxytech/dataplaneapi                    & 345     & 1024.0             \\
    HAProxy    & haproxytech/client-native                   & 133     & 4.0                \\
    HAProxy    & http-tests/cache-tests                      & 122     & 4.0                \\
    HAProxy    & haproxytech/config-parser                   & 82      & 4.0                \\
    HAProxy    & intel/workload-services-framework           & 55      & 128.0              \\
    HAProxy    & existentialcomics/kungFuChess               & 31      & 200.0              \\
    HAProxy    & xen0bit/muvr.xyz                            & 22      & 64.0               \\
    HAProxy    & zairo-korea/http\_request\_smuggling\_test  & 18      & 4.0                \\
    HAProxy    & haproxytech/haproxy-dev-lua-filters         & 5       & 200.0              \\

    \nginx     & Gallopsled/pwntools                         & 12518   & 1024.0             \\
    \nginx     & ShaneIsrael/fireshare                       & 757     & 500.0              \\
    \nginx     & huasenjio/huasenjio-compose                 & 560     & 51200.0            \\
    \nginx     & aldor007/mort                               & 513     & 73.0               \\
    \nginx     & PUGX/badge-poser                            & 477     & 500.0              \\
    \nginx     & brunobritodev/JPProject.IdentityServer4.SSO & 455     & 1024.0             \\
    \nginx     & dyc3/opentogethertube                       & 427     & 1000.0             \\
    \nginx     & DanWahlin/Angular-Docker-Microservices      & 220     & 3000.0             \\
    \nginx     & yuri-gushin/Roboo                           & 180     & 1000.0             \\
    \nginx     & PencilCode/pencilcode                       & 168     & 200.0              \\

    Squid      & bannedbook/fanqiang                         & 39502   & 5000.0             \\
    Squid      & vimagick/dockerfiles                        & 3181    & 100.0              \\
    Squid      & av/harbor                                   & 1576    & 100.0              \\
    Squid      & xjdrew/kone                                 & 705     & 10240.0            \\
    Squid      & diladele/squid-windows                      & 194     & 100.0              \\
    Squid      & salrashid123/squid\_proxy                   & 128     & 100.0              \\
    Squid      & 1265578519/PAC                              & 54      & 5000.0             \\
    Squid      & jacobproject/operation                      & 40      & 800.0              \\
    Squid      & CSCfi/ansible-role-squid                    & 9       & 100.0              \\
    Squid      & linuxgazette/lg                             & 5       & 20480.0            \\

    Varnish    & varnish/Varnish-Book                        & 353     & 256.0              \\
    Varnish    & wenerme/wener                               & 300     & 32768.0            \\
    Varnish    & aliuosio/mage2.docker                       & 58      & 2048.0             \\
    Varnish    & aqzt/docker-alpine                          & 53      & 100.0              \\
    Varnish    & localwiki/localwiki-backend-server          & 48      & 712.0              \\
    Varnish    & camptocamp/puppet-varnish                   & 36      & 256.0              \\
    Varnish    & roadiz/skeleton                             & 11      & 256.0              \\
    Varnish    & silentred/learning-path                     & 8       & 256.0              \\
    Varnish    & waldronlab/BugSigDB                         & 8       & 100.0              \\
    Varnish    & waldronlab/BugSigDB                         & 8       & 100.0              \\

    \bottomrule
    \end{tabular}
\end{table*}

Table~\ref{tab:app-cache-storage} lists a sample of the projects we detected on GitHub and analyzed for cache configurations. We present the top 10 projects for each caching proxy, sorted by popularity.

\end{document}